\documentclass[12pt]{article}

\usepackage[english]{babel}

\usepackage[a4paper,top=2.54cm,bottom=2.54cm,left=2.54cm,right=2.54cm,marginparwidth=0cm]{geometry}

\usepackage{amsmath}
\usepackage{graphicx}
\usepackage[colorlinks=true, allcolors=blue, hyperfootnotes=false]{hyperref}
\usepackage{setspace}
\usepackage[symbol, bottom]{footmisc}

\title{Rewinding the Void: How One Solution of Einstein's Field Equations Describes Both the Birth of a Universe and the End of Time}

\author{Charles W. Robson\footnote{charles.robson@tuni.fi (corresponding author)} \ and Marco Ornigotti\footnote{marco.ornigotti@tuni.fi}}

\date{}

\begin{document}

\maketitle

\vspace{-13mm}
\begin{center}
Laboratory of Photonics, Physics Unit, Tampere University, FI-33720 Finland

\end{center}
\vspace{5mm}

\begin{abstract}

The type D Kasner vacuum solution of general relativity is reviewed, highlighting the little-known, intriguing property that it can describe both an anisotropic cosmology and the spacetime deep within a black hole, these two descriptions linked by a reversal of the time-ordering of the solution. The flexible nature of solutions of Einstein's equations is emphasised, and a brief discussion of the arrow of time in modern physics is presented.

\end{abstract}

\newpage



\begin{center}
\emph{Time’s the king of men,}
\end{center}
\vspace{-15pt}
\begin{center}
\emph{He's both their parent, and he is their grave,}
\end{center}
\vspace{-15pt}
\begin{center}
\emph{And gives them what he will, not what they crave.}
\end{center}
\vspace{-10pt}
\begin{center}
from \emph{Pericles, Prince of Tyre} by William Shakespeare
\end{center}

\doublespacing

\section{The Supple Nature of General Relativity}

\label{sec:1}

Over the past century, general relativity, arguably the most beautiful of all physical theories, has been intensively explored. Countless hours have been devoted to forming a vast map of its highways and byways, leading to extraordinary results: black holes, gravitational waves, even entire universes can be described \cite{Stephani,Ehlers,MTW,Wald,Petrov_book,Belinski,Griffiths}. The theory has completely transformed our ideas about the fundamentals of nature and, despite the many attempts at finding inconsistencies or cracks in its edifice, it remains one of the most solid intellectual models of our time \cite{Carroll,LIGO,tests_review}.

One under-appreciated feature of general relativity is that a single solution of the Einstein field equations is capable of describing \emph{multiple} physical systems, though these systems may have vastly different features \cite{Griffiths}. This profound multivalency of solutions deserves to be better known, as it underlines the astonishing power and flexibility of the theory.


In this essay, we focus on one particularly intriguing example: the type D Kasner vacuum solution (which we call the ``Kasner solution'' from here on). It is able to describe both the cosmogony and expansion of a universe --- its birth and growth --- as well as the spacetime within a black hole \cite{Frolov,Frolov2,Hiscock,Ashtekar,Matyjasek,Universe_paper}. These two apparently distinct settings, the first moments of a universe and the last moments of an observer behind an event horizon, are described by an identical spacetime architecture --- one needs only to ``reverse'' the time coordinate in order to move between descriptions.

\section{Time in the Kasner Solution}

\label{sec:2}




Before considering its features, let us first look at the line element of the Kasner solution. It was discovered in the early twentieth century, not long after general relativity itself was first formulated \cite{Kasner1921,Harvey}. The Kasner solution has been interpreted as a homogeneous, anisotropic cosmological model, and its first fundamental form is given by \cite{Griffiths}
\begin{equation} \label{eq:Kasner}
ds^2 = -d\tau^2 + \tau^{-2/3} dz^2 + \tau^{4/3}(dx^2 + dy^2) .
\end{equation}

The coordinate $\tau$ defines a global time for a universe beginning with a big bang at $\tau=0$. We see from line element (\ref{eq:Kasner}) that a squeezing and stretching of space occurs as time progresses: at its inception, the spacetime is infinitely extended along the $z$ direction (a spacelike singularity), but as time advances there is a compression; on the other hand, the remaining spatial dimensions, $x$ and $y$, have no extent initially, both then growing along with $\tau$. In contrast to the well-known fate of an observer approaching the singularity of a black hole, this cosmological evolution can be termed ``de-spaghettification''.

Remarkably, the Kasner solution (\ref{eq:Kasner}) is also known to describe the spacetime deep within a Schwarzschild black hole, as mentioned in Sec. \ref{sec:1}. This can be demonstrated using simple coordinate transformations and approximations, as we now show.

The Schwarzschild solution of general relativity represents the unique spacetime region external to a source exhibiting only isotropic effects \cite{Lambourne}. It is the most elementary model of a black hole in physics, having only a single parameter: mass $M$. Starting from the Schwarzschild line element \cite{Carroll},
\begin{equation} \label{eq:Schwarz}
ds^2 = -\left( 1 - \frac{2M}{r} \right) dt^2 + \left( 1 - \frac{2M}{r} \right)^{-1} dr^2 + r^2 \left( d\theta^2 + \mathrm{sin}^2 \theta d\phi^2 \right) ,
\end{equation}
and taking the limits $r\ll 2M$ and $\theta\ll 1$, it is clear that the term $1-2M/r$ can be approximated by $-2M/r$ and that (by Taylor series) $\mathrm{sin}^2 \theta \approx \theta^2$. This yields
\begin{equation} \label{eq:metric_pos_M}
ds^2 \approx \frac{2M}{r} dt^2 - \frac{r}{2M} dr^2 + r^2 \left( d\rho^2 + \rho^2 d\phi^2 \right)
\end{equation}
after a relabelling $\theta \rightarrow \rho$ (this change of symbol just reflects the fact that the approximation given above has ``flattened'' the part of the metric proportional to $r^2$).

An infalling astronaut, after crossing the Rubicon of the black hole horizon, will inevitably \cite{Lambourne} arrive at the singularity $r=0$ and will feel the metric (\ref{eq:metric_pos_M}) as they approach it. The coordinate transformations $t \equiv \left( 3/4M \right)^{1/3} z, \quad r \equiv \left( 9M/2 \right)^{1/3} \tau^{2/3} \quad$ and $\rho e^{i\phi} \equiv \left( 2/9M \right)^{1/3} \left( x+iy \right)$
show that the line element (\ref{eq:metric_pos_M}) is in fact the Kasner solution (\ref{eq:Kasner}) in another guise \cite{Griffiths}.

How can one reconcile the fact that an observer falling into a black hole will, according to general relativity, be annihilated at the singularity with the statement that the same geometry describes the \emph{birth} of a spacetime? The key point is that the radial transformation given above, $r \equiv \left( 9M / 2 \right)^{1/3} \tau^{2/3}$, is also the relation between the radial coordinate of an observer falling\footnote[3]{The observer is assumed to fall purely radially, beginning at a very large distance away from the black hole.} into a Schwarzschild black hole and the \emph{countdown time} $\tau$ until that observer reaches the singularity \cite{MTW,Universe_paper,Lambourne}. The coordinate $r$, which has no immediate metrical significance (as with all coordinates in general relativity), is thereby related to a proper time $\tau$. This countdown time is defined as the proper time at which the singularity would be reached minus the proper time measured on the infalling observer's clock: $\tau = \tau_{\mathrm{sing}} - \tau_{\mathrm{observer}}$. As time $\tau$ ticks down for the doomed astronaut, the spacetime (\ref{eq:Kasner}) can be seen to extend along $z$ and contract circumferentially --- the observer is spaghettified \cite{Brief_history}.

The cosmology of the Kasner solution is then, in the above sense, a temporal reflection of the spacetime inside the black hole, the coordinate $\tau$ needing only to be inverted in direction in order to switch between the two descriptions. Two immensely different physical systems are then described using just one solution of Einstein's equations.

We remark finally that it is interesting that the presence of a past spacelike singularity, as one sees in the Kasner spacetime with $\tau$ flowing forwards, is a characteristic feature not only of big bangs but also of white holes, time-reversed black holes permitted by Einstein's theory \cite{Penrose}.

\section{Why Does Time Flow ``That Way''?}

We have seen above that a simple reversal in the time-ordering allows one to move between a description of a growing universe, with a singularity in the past, and that of the interior of a black hole, where a singularity lies in the future. But we know from experience that, in reality, time runs in only one direction. In our own universe, what is the reason that time flows one way and not the other?

Questions relating to the flow of time touch, inevitably, on the nature of time itself, an exceedingly deep and difficult problem that has been considered for millennia \cite{Rovelli}. Indeed, the concept of time is so central to our existence, so elemental, that it has been argued \cite{Huw} that the difficulty in our understanding of it is largely a result of it being, as it were, \emph{too central}: an objective viewpoint from ``outside'' of time --- where, for example, one could perform \emph{Gedankenexperimente} --- is extremely difficult to conceive and to express in language. Time's omnipresence also means that it is mostly taken for granted: as once put, ``[w]e are so familiar with the arrow of time that we are not often struck how odd it is'' \cite{Blundell}.

Nevertheless, general relativity has given us powerful tools to pull us out of our daily experience of time, allowing us (to a certain degree) to visualise the span of a universe as a four-dimensional block of spacetime, much the same way as one can see a three-dimensional object. The unique perspective of general relativity strips time of its preferential and enigmatic role by embedding it alongside the dimensions of space --- this union provides opportunities to tackle the arrow of time problem in an integrated manner.

Various approaches to the problem of time's arrow could, of course, be imagined, but here we touch solely on those attempts that have employed the second law of thermodynamics and entropy in some way, as these play an important role both in the physics of black holes and of cosmology.

An increase in entropy is an obvious marker of the passage of time, as is well known from thermodynamics \cite{Blundell}. The application of this concept to the entire universe, in order to explain the flow of time on a cosmological scale, naturally implies that the universe must have started in a low-entropy state, that a boundary condition is responsible for the subsequent flow. Roger Penrose has suggested \cite{Penrose_1979} that, although the beginning of our universe appears to have been highly entropic (in terms of matter), the Weyl curvature tensor, which encodes gravitational degrees of freedom, should vanish at any initial singularity and that this vanishing curvature may be related to a low ``entropy of gravity'' --- a geometric contribution to the entropy --- at the birth of our universe \cite{Ong,Tod}. Taking this into account, the overall entropy may then have increased since the Big Bang \cite{Lockwood}. Another proposal is that an arrow of time is tied to the expansion of the universe, but this idea becomes problematic when a universe with a contracting phase is considered \cite{Penrose_1979,Hawking85}.

Unsurprisingly, taking quantum mechanics into account complicates the matter profoundly. A quantum superposition of processes with opposite entropy variations is permitted --- in other words, a superposition of two opposing thermodynamic arrows of time can occur. Recent work has shown, however, that a definite thermodynamic arrow of time emerges in the quantum framework once a measurement of entropy production is performed \cite{Rubino}. It is conceivable that future experiments probing the emergence of an arrow of time using quantum measurement theory in a gravitational context will be very fruitful.

Whether or not the concept of entropy is a \emph{sine qua non} of explaining the nature and order of time on a fundamental level remains unclear, and a resolution may have to wait until a quantum theory of gravity is fully developed\footnote[4]{See Ref. \cite{QG} and references therein for a discussion of the problem of time in the context of quantum gravity.}.

\section{Reflections and Outlook} 

General relativity is a time-symmetric theory: if a solution of Einstein's field equations exists with one particular time-orientation, a reversal in the time-ordering can be performed whilst retaining it as a solution. In terms of the arrow of time, the parameter $\tau$ in the Kasner solution (\ref{eq:Kasner}) can either be chosen to go with, to ``guide'', the arrow or it can travel in the opposite direction, with the value of $\tau$ decreasing as past becomes future --- the former solution corresponds to the cosmological interpretation of the Kasner solution, the birth of a universe, whilst the latter accounts for its black hole interpretation, the singularity \emph{at the end of time}. It is important to stress that both of these pictures are valid and do not contradict each other whatsoever.

We hope that, in the future, taking the Kasner spacetime as an example, new solutions of Einstein's equations will be found (and old ones reappraised) providing more instances of single solutions capable of describing radically different zones. In particular, those solutions with clear physical interpretations for either time orientation may be useful for investigations into the one-directional flow of time.

\end{document}